\documentstyle[epsf,amsfonts,prl,aps,twocolumn]{revtex}

\begin{document}
\draft

\title{Strong Coupling Theory of Two Level Atoms in Periodic Fields}

\author{J. C. A. Barata and W. F. Wreszinski}

\address{
  Instituto de F\'{\i}sica.
  Universidade de S\~ao Paulo\\
  Caixa Postal 66 318\\
  05315 970. S\~ao Paulo. SP. Brasil \\
}

\maketitle 
\begin{abstract}
  We present a new convergent strong coupling expansion for two-level
  atoms in external periodic fields, free of secular terms. As a first
  application, we show that the coherent destruction of tunnelling is a
  third-order effect. We also present an exact treatment of the
  high-frequency region, and compare it with the theory of averaging.
  The qualitative frequency spectrum of the transition probability
  amplitude contains an effective Rabi frequency.
\end{abstract}



The advent of strong laser pulses has stimulated interest in
strong-coupling expansions in quantum optics and quantum
electrodynamics. Such expansions are also of considerable general
conceptual interest in several branches of physics. However,
particularly in the case of periodic and quasi-periodic perturbations,
the usual series, e.g., the Dyson series, are plagued by secular
terms, leading to a violation of unitarity when the expansion is
truncated at any order. In addition, small denominators appear in the
quasi-periodic case (see the discussion in the introduction in
\cite{qp}). These problems have been formally solved in a nice letter
by W. Scherer \cite{Scherer1} and in the papers which followed
\cite{Scherer2,Scherer3}. The main shortcoming in these works is that
convergence was not controlled, an admittedly difficult enterprise. By
writing an Ansatz in \underline{exponential} form, and
``renormalizing'' the exponential inductively, we were able to
eliminate completely the secular terms and to prove convergence in the
special case of a two-level atom subject to a periodic perturbation,
described by the Hamiltonian \cite{AutlerTownes}
\begin{equation}
      H_1 (t) \; = \; \epsilon \sigma_3 - f(t)\sigma_1 .
\label{hamiltoniano1}
\end{equation}
The corresponding Schr\"odinger equation is
\begin{equation}
     i \partial_t \Psi(t) \; = \; H_1(t) \Psi(t) ,
\label{primeira_equacao_de_Schroedinger}
\end{equation}
adopting $ \hbar = 1$ for simplicity.
Above $f(t)$ is of the form 
\begin{equation}
     f(t) \; = \; \sum_{n\in{\mathbb Z}} F_n e^{in\omega t} ,
\label{serie_de_f}
\end{equation}
with $ \overline{F_n} = F_{-n}$, since $f$ is real, and $\sigma_i $
are the Pauli matrices satisfying $ [\sigma_1, \; \sigma_2] =
2i\sigma_3$ plus cyclic permutations.  Assuming $F_n$ of order one,
the situation where $ \epsilon$ is ``small'' characterizes the strong
coupling domain.

It is convenient to perform a time-independent unitary rotation of $
\pi/2$ around the 2-axis in (\ref{hamiltoniano1}), replacing $ H_1(t)$
by 
\begin{equation}
     H_2 (t) \; = \; \epsilon \sigma_1 + f(t) \sigma_3 
\label{hamiltoniano2}
\end{equation}
and the Schr\"odinger equation by 
\begin{equation}
     i \partial_t \Phi(t) \; = \; H_2(t)\Phi(t) , 
\label{segunda_equacao_de_Schroedinger}
\end{equation}
with $ \Phi(t) = \exp(-i\pi \sigma_2/4)\Psi(t)$. 

The following result was proved in \cite{qp}. Let $ f$ be
continuously differentiable and $ g$ be a particular solution of the
generalized Riccati equation 
\begin{equation}
      g' -i g^2 - 2ifg + i\epsilon^2 \; = \; 0 .
\label{Riccati}
\end{equation}
Then the function $ \Phi: {{\mathbb R}} \to {{\mathbb C}}^2$ given by
\begin{equation}
     \Phi(t) \; = \; 
\left(
\begin{array}{c}
\phi_+(t) \\ \phi_-(t)
\end{array}
\right)
\; = \; 
U(t) \Phi(0) ,
\end{equation}
where
\begin{equation}
U(t)  \equiv 
\left(
\begin{array}{cc}
R(t)(1+ig_0 S(t)) & 
-i\epsilon R(t)S(t)
\\
   &
\\
-i\epsilon \overline{R(t)}\overline{S(t)} & 
\overline{R(t)}(1-i\overline{g_0}\overline{S(t)})
\end{array}
\right) ,
\label{solucao_de_U}
\end{equation}
with $ g_0 \equiv g(0)$,
\begin{equation}
     R(t) \; \equiv \; 
\exp
\left(
        -i \int_0^t (f(\tau) + g(\tau))\, d\tau
\right) 
\label{definicao_de_R}
\end{equation}
and
$ 
    S(t)  \equiv \int_0^t R(\tau)^{-2} \, d\tau
$, 
is a solution of the Schr\"odinger equation
(\ref{segunda_equacao_de_Schroedinger}) with initial value
$ 
\Phi(0) = \left(
\begin{array}{c}
\phi_+(0) \\ \phi_-(0)
\end{array}
\right)
$. 
A simple computation \cite{qp} shows that the components $ \phi_\pm$
of $ \Phi(t)$ satisfy a complex version of Hill's equation 
\begin{equation}
      \phi{_\pm}'' + (\pm if' + \epsilon^2 + f^2)\phi_{\pm} \; = \; 0.
\label{Hillqp}
\end{equation}
In \cite{qp} we attempted to solve (\ref{Hillqp}) using the Ansatz
$ 
  \phi(t) = \exp \left( -i \int_0^t (f(\tau) + g(\tau))d\tau \right)
$, 
from which it follows that $ g$ has to satisfy the generalized Riccati
equation (\ref{Riccati}). A similar idea was used by F. Bloch and A.
Siegert in \cite{BlochSiegert}. For $ \epsilon \equiv 0$ a solution of
(\ref{Riccati}) is given by $ \exp \left( -i \int_0^t f(\tau)d\tau
\right)$. Thus, in the above Ansatz we are searching for solutions in
terms of an ``effective external field'' of the form $f+g$, with $ g$
vanishing for $ \epsilon=0$.  It is thus natural to pose
\begin{equation}
    g(t) \; = \; \sum_{n=1}^{\infty} \epsilon^n G^{(n)} (t) ,
\label{seriedeg}
\end{equation}
where
\begin{equation}
     G^{(n)} (t) \; \equiv \; q(t)c_n (t)
\label{Gn}
\end{equation}
and
\begin{equation}
     q(t) \; \equiv \; \exp\left( i\int_0^t f(\tau)d\tau \right) .
\label{definicao_de_q}
\end{equation}
Inserting (\ref{seriedeg})-(\ref{Gn}) into (\ref{Riccati}) yields a
sequence of recursive equations for the coefficients $ c_n$, whose
solutions are
\begin{eqnarray}
 c_1 (t) & = & \alpha_1 \, q(t),  
 \label{S1} 
\\
 c_2 (t) & = & 
  q(t)  \, \left[ i \int_{0}^{t} 
        \left(\alpha_1^2 q(\tau)^2 - q(\tau)^{-2}  \right) d\tau + \alpha_2
          \right],      
  \label{S2} 
\\
 c_n (t) & = & 
         q(t)  \, \left[ i \left(\int_{0}^{t} \sum_{p=1}^{n-1}
              c_p(\tau) c_{n-p}(\tau) \, d\tau \right) + \alpha_n
              \right],    
  \label{S3} 
\end{eqnarray}
for $ n\geq 3$, where the $ \alpha_n$ are arbitrary integration
constants. The main point is that these constants may be chosen
inductively in order to cancel the secular terms. For instance, in
order to cancel the secular term in $ c_2$ in (\ref{S2}), the
integrand cannot contain a constant term, which equals the mean-value
term
\begin{equation}
M(q^2) \; \equiv \; \lim_{T\to\infty} \frac{1}{2T}\int_{-T}^T
q^2(\tau)\, d\tau \; \neq \; 0 .
\label{Mq2}
\end{equation}
Then it follows from (\ref{S2}) that one must require
\begin{equation}
M(\alpha_1^2 q^2 - q^{-2}) \; = \; 0 \; \Longrightarrow \; \alpha_1^2 \; = \;
\frac{\overline{M(q^2)}}{M(q^2)} .
\label{alpha_1}
\end{equation}
It was proved in \cite{qp} that one may proceed in this way and
establish the absence of secular terms of any order. Similar results
are valid if (\ref{Mq2}) is not satisfied. 

In the quasi-periodic case we were not able to show convergence in
(\ref{seriedeg}), and, in fact, it is not expected \cite{qp}. Hence,
(\ref{seriedeg}) is to be viewed as a formal power series. In the
periodic case (\ref{serie_de_f}) much stronger results are possible,
as we now discuss. 

Let $ G_m^{(n)}$, $ C_m^{(n)}$, $Q_m$ and $ Q_m^{(2)}$ denote the
Fourier coefficients of $G^{(n)}(t)$, $c_n(t)$ (given in
(\ref{seriedeg})-(\ref{Gn})), $q(t)$ (given in (\ref{definicao_de_q}))
and $q^2(t)$, respectively, defined as in (\ref{serie_de_f}).  Due to
the multiplication by $ q(t)$ in (\ref{Gn}) the $ G_m^{(n)} $ are
given by convolutions
\begin{equation}
  G_m^{(n)} \; = \; \sum_{l=-\infty}^{\infty} Q_{m-l}C_l^{(n)} .
\label{convolucao_1}
\end{equation}
The $C_l^{(n)}$, the Fourier components of $c_n(t)$, have, by
(\ref{S1})-(\ref{S3}), explicit expressions in terms of the $ Q_m $
and $Q_m^{(2)} $, for instance, if (\ref{Mq2}) holds and $ \alpha_1$
is given by (\ref{alpha_1}), 
\begin{eqnarray}
C_{m}^{(1)} & = & \alpha_1 Q_{m} ,
\label{PerCFourier1}
\\
C_{m}^{(2)} & = & 
       \sum_{n =\infty\atop n \neq 0}^{\infty}
       \frac{\left( \alpha_1^2 Q^{(2)}_{n} -
         \overline{Q^{(2)}_{-n}} \right)}{n  \omega}
     \left[
          Q_{m - n} - \frac{Q_{m} Q_{ -n}^{(2)}}{
                             Q_{0}^{(2)} }
       \right] ,
\label{PerCFourier2}
\end{eqnarray}
Note that, above, $Q_0^{(2)} = M(q^2) \neq 0$. The solution for the
case $ M(q^2) = 0 $ is found in \cite{qp,pp}.  Finally, let
\begin{equation}
     \Omega \; = \; \Omega (\epsilon) \; \equiv \; F_0 + G_0 (\epsilon) ,
\label{definicao_de_Omega}
\end{equation}
\begin{equation}
H_n  \equiv  \left\{ \begin{array}{cl}
                            [F_n + G_n(\epsilon)](n\omega)^{-1} , 
                                       & \mbox{for } n\neq 0 \\
                                       & \\
                             0,        & \mbox{for } n=0
               \end{array}
         \right. ,
\label{definicao_de_Hn}
\end{equation} 
and
$ \displaystyle 
     \gamma_f(\epsilon)  \equiv   i \sum_{m\in {\mathbb Z} } \, H_m 
$, 
with
$ \displaystyle 
      G_m (\epsilon)  \equiv  \sum_{n=1}^\infty G_{m}^{(n)} \epsilon^n 
$. 
Then $ R(t)$, in (\ref{definicao_de_R}), is given by 
\begin{equation}
    R(t) \; = \;  e^{-i\gamma_f(\epsilon)}\;e^{-i \Omega t}\;
                   {\cal R}(\infty, \; t) ,
\label{isuybvgpweruewp}
\end{equation}
with
$ \displaystyle 
{\cal R}(N, \; t) \equiv \exp \left( -\sum_{n= -N}^{N} H_n e^{in\omega
    t} \right)
$. 
By (\ref{solucao_de_U}), the complete wave function is known once
(\ref{isuybvgpweruewp}) is given; (\ref{solucao_de_U}) and
(\ref{isuybvgpweruewp}) also show that the wave-function is of the
\underline{Floquet form}, with secular frequencies $\pm \Omega$. 

In reference \cite{pp} we have proven the following result: {\it for $
  f$ periodic the $\epsilon$-expansion (\ref{seriedeg}) has a nonzero
  radius of convergence}. Our estimate for this radius is not optimal
and we refrain from quoting it here, but we remark that the expansion
does converge for high frequencies, i.e., $ \omega \gg \epsilon $, a
condition that we assume in the following.

We now consider in (\ref{serie_de_f}) the special case
\begin{equation}
     F_n \; = \; \frac{1}{2}(\delta_{n, \; 1} +\delta_{n, \; -1 }) ,
\label{Fn_e_delta}
\end{equation}
corresponding to
$ 
      f(t)  =  \cos (\omega t) 
$. 
For this case
\begin{equation}
      Q_m \; = \; J_m \left( \frac{1}{\omega}\right)
\; \mbox{ and }  \; 
Q_m^{(2)} \; = \; J_m \left( \frac{2}{\omega}\right) .
\label{QeJ}
\end{equation}
By (\ref{alpha_1}), 
$ 
    \alpha_1 =  1 
$. 

The transition amplitude $ A_{21}$ from the lowest energy atomic state
$ 
\left(
{0 \atop 1}
\right)
$ 
of (\ref{hamiltoniano1}) to the upper level 
$ 
\left(
{1 \atop 0}
\right)
$ 
is
\begin{equation}
      A_{21}(t) \; = \; \left( \psi_1, \; U(t) \psi_2 \right) ,
\end{equation}
where $U(t)$ is given by (\ref{solucao_de_U}) and
$ 
\psi_2  =  \frac{1}{\sqrt{2}}
\left(
{1 \atop -1}
\right)
$, 
$  
\psi_1  =  \frac{1}{\sqrt{2}}
\left(
{1 \atop 1}
\right)
$ 
are the corresponding eigenstates of the rotated Hamiltonian $H_2$,
given by (\ref{hamiltoniano2}).  The \underline{tunnelling} amplitude
corresponds to the transition probability
\begin{equation}
\tilde{A}_{21} (t ) \; = \; 
\left( \tilde{\psi}_1, \; U(t) \tilde{\psi}_2 \right) ,
\end{equation}
where 
$ 
\tilde{\psi}_2 = \left( {0 \atop 1} \right)
$, 
$  
\tilde{\psi}_1 = \left( {1 \atop 0} \right)
$. 
The latter represent the localized states (eigenstates of the field
term, proportional to $ \sigma_3 $, in (\ref{hamiltoniano2})), and $
\tilde{A}_{21} = 0$ means absence of tunnelling between these states.
Indeed, (\ref{hamiltoniano2}) is a semi-classical approximation to the
spin-Boson system treated in \cite{SpohnDuemcke}.  In the full
quantized case, considered in \cite{SpohnDuemcke}, $ \tilde{\psi}_1$
and $ \tilde{\psi}_2$ differ macroscopically because they are dressed
by photon clouds, and for $ \epsilon $ sufficiently small there is always
localization, i.e., no tunnelling. This is not the case here, as we
shall see.

\begin{figure}[h!]
\leavevmode
\vbox{
 \centerline{
  \epsfbox[130 304 484 473]{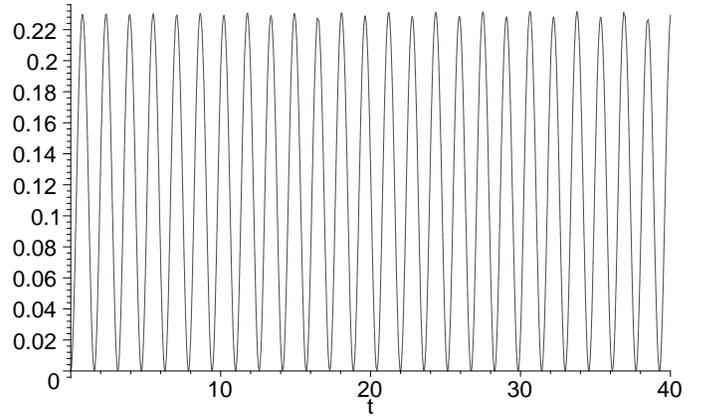}}
 }
 \caption{The amplitude $|A_{21}(t)|^2$. 
          Here $\epsilon=0.01$ and $\omega =2 $.}
\label{fig1}
\end{figure}

\begin{figure}[h!]
  \leavevmode
\vbox{
 \centerline{
  \epsfbox[130 304 484 465]{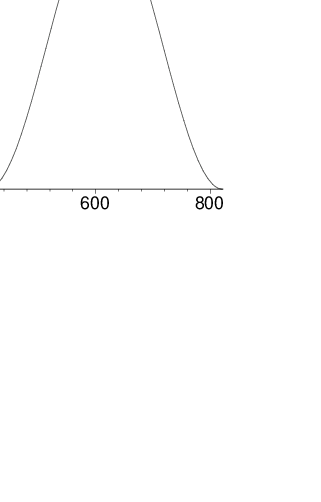} 
 }
}
\caption{The amplitude $|\tilde{A}_{21}(t)|^2$. 
         Here $\epsilon=0.01$ and $\omega =2 $.}
\label{fig2}
\end{figure}

Fig.\ \ref{fig1} shows the exact result for $|A_{21}(t)|^2$ to fifth
order in $ \epsilon$ for about $12$ cycles of $ \omega$.  We see clearly
the domination of the external frequency $ \omega$, in agreement with
the theory of averaging. Eq.\ (\ref{segunda_equacao_de_Schroedinger})
may be transformed to
\begin{equation}
\partial_t \tilde{\psi} \; = \; \epsilon f(t, \; \tilde{\psi})
\label{jdsld_1}
\end{equation} 
with 
$ 
\tilde{\psi} = \exp \left( i\int_0^t f(\tau)d\tau \right)\psi
$, 
and 
\begin{equation}
f(t, \; \tilde{\psi}) \; = \; 
\left(
\begin{array}{cc}
0 & e^{2i\sin(\omega t)/\omega}
\\
e^{-2i\sin(\omega t)/\omega} & 0
\end{array}
\right) \tilde{\psi} .
\label{jdsld_2}
\end{equation}
By (\ref{jdsld_1}) and (\ref{jdsld_2}), the averaged equation
$\partial_t \tilde{\psi_0}  =\epsilon f^{(0)}(\tilde{\psi_0})$
with 
$ 
   f^{(0)}(\tilde{\psi_0}) = \frac{1}{T}\int_0^T f(t, \, \tilde{\psi_0})
$ 
and $ T = 2\pi/\omega$, is
\begin{equation}
i \partial_t \tilde{\psi_0} \; = \; \epsilon J_0\left(\frac{2}{\omega} \right) 
\sigma_1 \tilde{\psi_0}
\end{equation}
and a well known theorem \cite{Verhulst} yields
$ 
\left| \tilde{\psi}(t) - \tilde{\psi_0}(t)\right|
=
O(\epsilon/\omega)
$ 
on the time scale $1/\epsilon$. Hence $A_{21}$ is close to 
$$
\left( \psi_1 , \; \exp\left( -i\epsilon J_0
    \left(\frac{2}{\omega}\right)\sigma_1 t\right) \exp\left( i
    \frac{\sin(\omega t)}{\omega}\sigma_3 \right) \psi_2 \right)
\; = \; 
$$
$$
i\exp\left(-i\epsilon J_0\left(\frac{2}{\omega}\right)t \right)
\sin \left(\frac{\sin(\omega t)}{\omega} \right) .
$$
Since 
$ \displaystyle 
\sin \left(\frac{\sin(\omega t)}{\omega} \right)  = 
2 \sum_{k=0}^{\infty}
J_{2k + 1} \left( \frac{1}{\omega} \right)
\sin \left[ (2k+1)\omega t\right] 
$, 
we see that in this case the spectrum is dominated by the harmonics of
the frequency $\omega$ of the external field, in agreement with Fig.\ 
\ref{fig1}. Notice, however, that, while averaging is applicable to
times up to $ O(1/\epsilon)$, the exact theory is applicable to
\underline{all} times. Applying the averaging theory to
$\tilde{A}_{21} $, we are led to the matrix element
$$
\left(
\exp\left( i \frac{\sin(\omega t)}{\omega}\sigma_3 \right)
\tilde{\psi}_1
, \; 
\exp\left(-i\epsilon J_0\left(\chi \right)\sigma_1 t \right)
\tilde{\psi}_2
\right)
 =  
$$
$$
-ie^{i \frac{\sin(\omega t)}{\omega}}
\sin
\left(
\epsilon J_0 \left( \chi \right) t
\right)
\, \simeq \, 
-i J_0 \left( \frac{\chi}{2} \right)
\sin
\left(
\epsilon J_0 \left( \chi \right) t
\right) ,
$$
with $ \chi \equiv 2/\omega$.  This result agrees with
(\ref{isuybvgpweruewp}) and Fig.\ \ref{fig2}, which shows the exact
result for $|\tilde{A}_{21}(t)|^2$ to fifth order in $ \epsilon$ for $t$
from $ 0$ to $2\pi/\Omega $.  Fig.\ \ref{fig2} shows that $
\Omega(\epsilon) $, the secular frequency given by
(\ref{definicao_de_Omega}), dominates in this case.  There,
$\Omega(\epsilon) \simeq 7.6 \, 10^{-3}$ for the values of $\epsilon$ and
$\omega$ chosen.

Notice that by (\ref{convolucao_1}), (\ref{PerCFourier1}),
(\ref{definicao_de_Omega}) and (\ref{QeJ})
$ 
\Omega (\epsilon) \simeq \epsilon J_0 (\chi ) 
$, 
to first order in $ \epsilon$. Thus, the first order contribution
approaches zero if $ \chi$ approaches one of the zeros of the Bessel
function $ J_0$.  The second order contribution to $\Omega$ is
$ \displaystyle 
\epsilon^2 \sum_{l\in {\mathbb Z}} Q_{-l}C_{l}^{(2)}
$, 
and is \underline{identically} zero, as one sees using
(\ref{PerCFourier2}). The third order contribution to $\Omega$ is
$$
-\frac{2\epsilon^3}{\omega^2}
\sum_{
   n_1, \; n_2 = -\infty 
}^{\infty}
 \frac{J_{2n_1+1}(\chi) J_{2n_2 +1}(\chi)
      J_{-2(n_1+n_2 +1)}(\chi)}{(2n_1+1)(2n_2+1)}
$$
and is non-zero if $ \chi$ coincides with one of the zeros of $ J_0$.
Hence, when $ 2/\omega$ approaches one of the zeros of the Bessel
function $ J_0$ the lowest non-vanishing contribution to $ \Omega$ is
of third order in $\epsilon$ and, hence, rather small. This means that for
such values of $\omega$ the tunnelling is very heavily, although not
exactly, suppressed.

Hampering and destruction of tunnelling have been studied in
\cite{GrossmanDittrichJungHanggi,Kayanuma} for particles, and in
\cite{vanHemmemSuto,vanHemmemWreszinski} for spins. The latter use the
method of averaging, but we emphasize that in the case treated above,
$ \omega \gg \epsilon $ is satisfied for $ \epsilon$ sufficiently small, and
thus the result is exact, i.e., valid for all times. In addition, the
features regarding the order of the expansion are new.

\begin{figure}[h!]
  \leavevmode
\vbox{
 \centerline{
  \epsfbox[130 304 484 465]{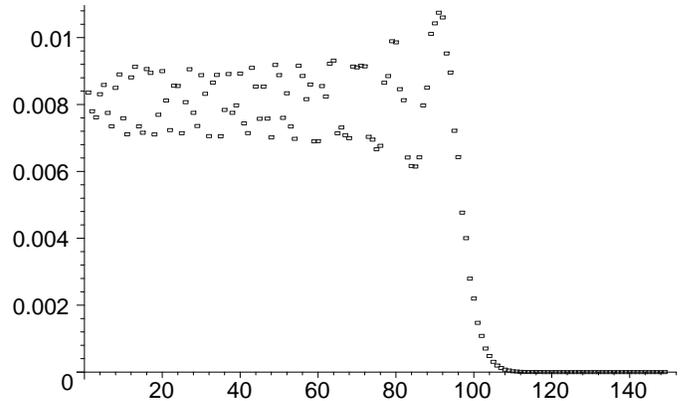} 
 }
}
\caption{The quantity ${\cal E}(N, \; t) $ as a function of $ N$. }
\label{fig3}
\end{figure}

At resonance $ \omega = 2\epsilon$ we are not able to prove that the
expansion converges. It is, nevertheless, a well-defined formal
expansion, in contrast to strong-coupling approximations of Keldish
type, which are beset with difficulties (see, e.g. \cite{BDM} and
references given there). Moreover, as we shall show, it includes
interesting effects of dressing of the atoms by the photon field (in
the semi-classical approximation) which yields the external field
Floquet description, rigorously justified in \cite{GMDJ}. Such effects
appear in the rotating-wave-approximation (RWA) in the form of a Rabi
frequency (see, e.g.  \cite{GMDJ}), but the present model is not close
to RWA, since the rotating and counter-rotating terms in
(\ref{hamiltoniano1}) are of the same order of magnitude. Moreover,
the RWA is not justified for large coupling, but the
\underline{solution} of (\ref{segunda_equacao_de_Schroedinger}) might
have some similarity to the solution obtained when the RWA is
performed. If so, the effective frequency of oscillation of $ A_{21}$
would not differ much from the Rabi frequency (see \cite{GMDJ})
\begin{equation}
     \Omega_R \; = \; \left[(\omega -2\epsilon)^2 + 4\right]^{1/2} \simeq
     2 
\label{frequencia_ed_Rabi}
\end{equation}
for $ \omega \simeq 2\epsilon$ (or, in general, for $ \omega = O(\epsilon)$ and
$ \epsilon$ small). Indeed, $ \Omega_R$ makes its appearance in
(\ref{isuybvgpweruewp}) in a most interesting way: by
(\ref{convolucao_1})-(\ref{PerCFourier1})-(\ref{definicao_de_Hn}) and
(\ref{QeJ}), $ H_n$ in $ {\cal R}(\infty, \, t)$ equals, to first order in $
\epsilon$,
\begin{equation}
\frac{F_n}{n\omega} + \epsilon \frac{J_n \left( \frac{2}{\omega}
  \right)}{n \omega},
\label{wjfdjvhps}
\end{equation}
with $F_n$ given by (\ref{Fn_e_delta}). The greatest contributions of
(\ref{wjfdjvhps}) arises for small $ n$ (due to the factor $n^{-1}$)
and when the argument of the Bessel function equals its order, i.e., $
n = 2/\omega$, and the corresponding frequency in
(\ref{isuybvgpweruewp}) is $ n\omega = \frac{2}{\omega}\omega =2$,
which compares well with (\ref{frequencia_ed_Rabi}).  In Fig.\ 
\ref{fig3} we show this last effect for $ {\cal R}(N, \; t) $.  We
considered the quantity
$ 
{\cal E}(N, \; t) \equiv \left| \frac{{\cal R}(N, \; t) }{{\cal R}(\infty,
    \; t) } -1 \right|
$ 
which measures the error made by including in (\ref{isuybvgpweruewp})
only the first $N$ terms of the sum involving $H_n$ in ${\cal R}(\infty, \; t)
$.  In Fig.\ \ref{fig3} we considered the resonant case with $ \omega
= 2 \epsilon = 2\, 10^{-2}$ and $t =0.7\pi/\Omega $. The qualitative
behaviour is the same for other values of $t$.  We see from Fig.\ 
\ref{fig3} that mainly only small $ n$ and $n$ around $2/\omega$
contribute.  The effect of adding in the second order contribution is
negligible for the range of values of $ \epsilon$ considered.

In conclusion, the new strong-coupling expansion allows considerable
insight into both the high-frequency and resonance regimes, and
yields an interesting unexpected result for the coherent destruction
of tunnelling.

\acknowledgments

We would like to thank Dr.\ A.\ Sacchetti for a most valuable suggestion
regarding the coherent destruction of tunnelling.
We are also grateful to CNPq for partial financial support.




\end{document}